\newcommand{\beq}{\begin{equation}}
\newcommand{\eeq}{\end{equation}}
\newcommand{\bea}{\begin{eqnarray}}
\newcommand{\eea}{\end{eqnarray}}
\newcommand{\bear}{\begin{eqnarray*}}
\newcommand{\eear}{\end{eqnarray*}}

\documentclass[12pt]{article}
\begin{document}

\title{Supersymmetry on Jacobstahl lattices}
\author{Francisco~C.~Alcaraz$^{\rm a}$ and  Vladimir~Rittenberg$^{\rm b}$\\
\small \it $^{\rm a}$
Universidade de S\~ao Paulo,
Instituto de F\'{\i}sica de S\~ao Carlos, \\
\small \it C.P. 369,13560-590, S\~ao Carlos, SP, Brazil \\
\small \it $^{\rm b}$ Physikalisches Institut, Universit\"at Bonn, \\
\small \it Nussallee 12, D-5300 Bonn 1, Germany}

\date{}

\maketitle

\begin{abstract}
It is shown that the construction of Yang and Fendley  (2004 {\it J. Phys. A: Math. 
Gen. {\bf 37}} 8937) to obtain 
supersymmetric systems, leads not to the open XXZ chain with anisotropy 
$\Delta =-\frac{1}{2}$ 
but to systems having dimensions given by Jacobstahl sequences. 
For each system the ground state is unique. The 
continuum limit of the spectra of the Jacobstahl systems coincide,  up to 
degeneracies, with that of the $U_q(sl(2))$ invariant XXZ chain for 
$q=\exp (i\pi/3)$.
The relation between the Jacobstahl systems and the open XXZ chain is
explained. 
\end{abstract}

\vskip 1em

\noindent 
 Yang and Fendley \cite{YF} have given a construction to obtain 
 supersymmetric systems given by Hamiltonians $H$ obeying $N=2$ supersymmetry:
\beq
\label{eq1}
\{Q,Q^{\dag}\} = H, \,\,\,\,\,\,\ Q^2 =0,\,\,\,\,
\eeq
\beq
\label{eq2}
[F,Q]=-Q.
\eeq
 Here $Q$ and $Q^+$ are the supercharges.
 Their results imply that a special combination
 \beq \label{f1}
 H = \bigoplus_{L=1}^{\infty} H_{XXZ}^{(L)}
 \eeq
 of the  $L$ site XXZ open quantum chain, at anisotropy 
 $\Delta = -1/2$, which in terms 
 of the standard Pauli matrices is given by
 \beq
 \label{eq3}
 H_{ XXZ}^{(L)} = -\frac{1}{2} \sum_{i=1}^{L} \left( 
 \sigma_i^x \sigma_{i+1}^x +\sigma_i^y \sigma_{i+1}^y - 
 \frac{1}{2} \sigma_i^z \sigma_{i+1}^z \right) - \frac{1}{4}(\sigma_1^z + 
 \sigma_L^z) + \frac{3L-1}{4}
 \eeq
 is supersymmetric 
  and  therefore the ground-state energy is zero. The 
  last fact is correct but can be proven by other methods \cite{NGR}. 
  However for  the Hamiltonian (\ref{f1}) 
  one cannot derive a continuum limit or even 
  compare it with experiments. In this Letter we are going to show that 
  a another use of the supercharges defined by Yang and Fendley \cite{YF} can 
  lead to the construction of finite-dimension quantum spins with a well
  defined thermodynamical limit ($L \rightarrow \infty$). 
 We also give a possible explanation of
 the observation \cite{N-J-B} that the 
 ground-state wavefunction of $H_{XXZ}^{(L)}$ 
 is given by positive integers.   

 Instead of defining the $Q$ operator given in (\ref{eq1})-(\ref{eq2}) in terms 
 of fermionic operators, as done in \cite{YF} we give their matrix elements 
 on an appropriate  representation of the vector space. This operator, 
 that we denote by $Q^{(L,L+1)}$,  acts on a Hilbert space of dimension 
 $2^L \oplus 2^{L+1}$ spanned by the vector basis $\{|v_L>\} 
 \oplus\{ |v_{L+1}>\}$, where $\{|v_L>\} = \{|s_1,\ldots,s_L>\}$, 
 $\{|v_{L+1}>\} = \{|s'_1,\ldots,s'_{L+1}>\}$, $s_i, s'_j = \pm $ 
 , is the standard $\sigma_z-$basis, namely,
 \bea
 \label{eq4}
 &&Q^{(L,L+1)} = \sum_{j=1}^L Q_j^{(L,L+1)},\nonumber \\
 &&Q_j^{(L,L+1)} |s_1',\ldots,s_j',\ldots,s_{L+1}'> =  0, \; \; 
 Q_j^{(L,L+1)} |s_1,\ldots,s_j,\ldots,s_L> = \nonumber \\ 
 && (-)^{j-i}  |s_1,\ldots,s_{j-1},+,+,s_{j+1},\ldots,s_L> 
 \delta_{s_j,-},
 \eea
 where ($j=1,\ldots,L$). 
 It is immediate from (\ref{eq4}) to see  that $Q_i^{(L,L+1)}   
 Q_j^{(L,L+1)}=0$, 
 implying  $(Q^{(L,L+1)})^2 =0$.
 We can define non-local $L$-site quantum chains $H_1^{(L)}$ and 
 $H_2^{(L)}$ acting on the vector space spanned by $\{|v_L>\}$ of dimension 
 $2^L$ by 
 \bea
 \label{eq5}
 Q^{(L,L+1)} {Q^{(L,L+1)}}^{\dag} &=& O^{(L)}\oplus H_2^{(L+1)}, \nonumber \\
 {Q^{(L,L+1)}}^{\dag} Q^{(L,L+1)} &=&  H_1^{(L)}\oplus O^{(L+1)} 
\eea
where $O^{L'}$ ($L'=L,L+1$)
is the matrix with zero elements in the space of dimension 
$2^{L'}$ spanned by $\{|v_{L'}>\}$. We can verify the following properties 
of $H_1^{(L)}$ and $H_2^{(L)}$: 
\bea 
\label{eq6}
&& {H_1^{(L)}}^{\dag} = H_1^{(L)}, \,\, {H_2^{(L)}}^{\dag} = H_2^{(L)}, \,\,\, 
 [S^z,H_1^{(L)}] =[S^z,H_2^{(L)}] =0,  \nonumber \\ 
&&  S^z|s_1,\ldots,s_L> = 
(\sum_{i=1}^L s_i)|s_1,\ldots,s_L>. 
\eea

We can also verify that  that   
\beq 
\label{eq6'}
H_s^{(L+1)} = \{Q^{(L,L+1)},
{Q^{(L,L+1)}}^{\dag}\} = H_1^{(L)} \oplus H_2^{(L+1)} \neq 
H_{XXZ}^{(L)}.
\eeq
and consequently the $L$ site XXZ open chain is not supersymmetric.

The correct relation  
among these quantum chains with the $H_{XXZ}^{(L)}$ is given by
\beq 
\label{eq7}
H_{XXZ}^{(L)} = H_1^{(L)} + H_2^{(L)}.
\eeq
Moreover we can also verify that 
\beq 
\label{eq8}
H_1^{(L)} H_2^{(L)} = H_2^{(L)} H_1^{(L)}  = 0,
\eeq
that imply that $H_1^{(L)}$, $H_2^{(L)}$ and $H_{XXZ}^{(L)}$
 share the same eigenvectors, and the non-zero eigenvalues of 
$H_1^{(L)}$ or $H_2^{(L)}$ are the same as those of $H_{XXZ}^{(L)}$ .
 For general values of $L$ the Hamiltonian 
$H_1^{(L)}$ obtained from (\ref{eq5}) is given by 
\bea 
\label{eq9}
H_1^{(L)}& =& - \frac{1}{2} \sum_{i=1}^{L} (\sigma_i^z -1) - 
\sum_{i=1}^{L-1} (\sigma_i^+\sigma_{i+1}^- + \sigma_i^-\sigma_{i+1}^+) \nonumber
\\
&& +\frac{1}{8} \sum_{i=1}^{L-2}\sum_{k=i+2}^{L} (-)^{k-i}\sigma_i^-
\sigma_k^+ A_{i,k}(\sigma_i^z+1)(\sigma_{i+1}^z+1)(1-\sigma_k^z) \nonumber \\
&& +\frac{1}{8} \sum_{i=3}^{L}\sum_{k=1}^{i-2} (-)^{k-i}\sigma_i^-
\sigma_k^+ A_{k,i}^{\dag} (\sigma_{i-1}^z+1)(\sigma_{i}^z+1)(1-\sigma_k^z),
 \eea
where  
\bea
\label{eq9p}
A_{i,k} = \cases{ 0 & if $ k<i+2$ \cr
1 & if $k =i+2$ \cr
(\frac{1}{2}\vec{\sigma}_{k-2}\vec{\sigma}_{k-1}+\frac{1}{2})\cdots
(\frac{1}{2}\vec{\sigma}_{i+2}\vec{\sigma}_{i+3}+\frac{1}{2})
(\frac{1}{2}\vec{\sigma}_{i+1}\vec{\sigma}_{i+2}+\frac{1}{2})& if $ k>i+2$ \cr}
\nonumber
\eea
 and $\vec{\sigma} = (\sigma^x,\sigma^y,\sigma^z)$,  $\sigma^{\pm}= (\sigma^x \pm \sigma^y)/2$ and $[L/2] = \mbox{Int}(L/2)$.
 The quantum chain $H_2^{(L)}$ is obtained from $H_1^{(L)}$ and 
 $H_{XXZ}^{(L)}$ by using (\ref{eq3}) and (\ref{eq7}) . Some examples for 
 $L=2$ and $L=3$ are:
 \bea 
 \label{eq10}
 H_1^{(2)} &=& -\frac{1}{2}(\sigma_1^x\sigma_2^x + \sigma_1^y\sigma_2^y) 
 - \frac{1}{2}(\sigma_1^z+ \sigma_2^z) + 1, \nonumber \\
 H_2^{(2)} &=& \frac{1}{4}\sigma_1^z\sigma_2^z + \frac{1}{4}(\sigma_1^z+ 
 \sigma_2^z) + \frac{1}{4},
 \eea
 \bea
 \label{eq11}
 H_1^{(3)} &=& -\frac{1}{2}\left[(
 \sigma_1^x\sigma_2^x + \sigma_1^y\sigma_2^y + \sigma_2^x\sigma_3^x + 
 \sigma_2^y\sigma_3^y) - \frac{1}{2}
 (\sigma_1^x\sigma_3^x + \sigma_1^y\sigma_3^y)
 (1 +\sigma_2^z)  \nonumber \right. \\
 && + \left.  \sigma_1^z + \sigma_2^z+ \sigma_3^z  -3 \right]  \nonumber \\
 H_2^{(3)}&=& -\frac{1}{4}\left[ (\sigma_1^x\sigma_3^x+\sigma_1^y\sigma_3^y) 
 (1+\sigma_2^z) -(\sigma_1^z\sigma_2^z+\sigma_2^z\sigma_3^z)\right.
 \nonumber \\
 &&-\left. (\sigma_1^z+2\sigma_2^z+\sigma_3^z) -2\right ] .
 \eea
 
 Notice that in order to find the eigenvalues and eigenfunctions of
$H_1^{(L)}$ and $H_2^{(L)}$ one has to use the eigenfunctions of 
$H_{XXZ}^{(L)}$
which 
can be obtained  using the Bethe Ansatz.
 Of the $2^L$ eigenvalues of $H_{XXZ}^{(L)}$,
 \beq 
 \label{eq6''}
 \frac{1}{3}\left[2^L -\frac{[3-(-1)^L]}{2}\right]
 \eeq
can be found in $H_2^{(L)}$ and the remaining 
 \beq 
 \label{eq6'''}
 \frac{1}{3}\left[ 2^{L+1} +\frac{[3-(-1)^L]}{2} \right]
 \eeq
 in $H_1^{(L)}$.   The ground-state energy being included in this last set.
All the eigenvalues of $H_1^{(L)}$ and $H_2^{(L)}$ not belonging to 
the sets (\ref{eq6''}) and (\ref{eq6'''}) are equal to zero.

 The quantum chain $H_{XXZ}^{(L)}$ although having a zero-energy ground 
 state and  all the eigenlevels real and positive numbers is not 
 supersymmetric. 
 The Hamiltonian $H_s^{(L+1)}$  with $3\times2^L$ states given by 
 (\ref{eq6'}) is 
supersymmetric. The supercharges connect states with $S^z=m$ in the $2^L$
vector space with states with $S^z=m+3$ in the $2^{(L+1)}$ vector space. The 
non-zero energies appear in doublets (some of them degenerate) and the
zero energy is highly degenerate. The degeneracies of the
zero energy level can be computed using eqs.(\ref{eq6''}) and (\ref{eq6'''}).
 Can we define supersymmetric systems with an unique ground-state and the 
other energy levels coinciding with those of $H_s^{(L)}$? The answer is yes but 
the path is long. One starts with $L=2$ and take the 2 states corresponding
to $S^z=0$ ($|-+>$ and $|+->$) and one state with $S^z=-2$ ($|-->$) 
to which we 
apply $Q^{(2,3)}$ defined by (\ref{eq4}). One gets the 2 states $|+++>$ and 
($|++->$-$|-++>$). Using this definition of $Q^{(2,3)}$ in 
this subspace only,  
 one obtains a supersymmetric system with 5 states defined by the 
Hamiltonian $H_J^{(3)}$. What we have done is to truncate  the vector space of 
$H_1^{(2)}\oplus H_2^{(3)}$. 
 We denote by $H_{1,t}^{(2)}$ and $H_{2,t}^{(3)}$ the Hamiltonians 
acting in the truncated  spaces. Obviously
\beq 
\label{eq6''''}
H_J^{(3)} = H_{1,t}^{(2)} \oplus H_{2,t}^{(3)}.
\eeq
In order to obtain the truncated  vector space in which $H_{1,t}^{(3)}$ acts, one 
has to take the vector space which is orthogonal to the one which 
$H_{2,t}^{(3)}$ acted. This is a 6-dimensional vector space: $S^z=-3$ (1 state),
$S^z=-1$ (3 states) and $S^z=1$ (2 states: ($|++->+|-++>$) and $|+-+>$). One 
can proceed further. (A look at Appendix A of Ref. \cite{NGRP} might help the
reader to follow the steps).  One uses $Q^{(3,4)}$ applied to the 6 states to find the
vector space in which $H_{2,t}^{(4)}$ acts and find that supersymmetric 
Hamiltonian $H_J^{(4)}$ which acts in an 11 dimensional space etc... Applying 
consistently this procedure, we find that the supersymmetric 
Hamiltonian
\beq 
\label{eq6v}
H_J^{(L)} = H_{1,t}^{(L-1)} \oplus H_{2,t}^{(L)} 
\eeq
acts in a vector space of dimension
\beq
\label{eq6vv}
\frac{1}{3}\left[ 2^{L+1} + (-1)^L \right] .
\eeq
The numbers obtained using eq.(\ref{eq6vv}) are called Jacobstahl 
numbers and they 
have interesting combinatorial interpretations \cite{heubach}.
 One can use now the results of Ref.\cite{NGRP} (eqs.(3.5) and (3.6)) to obtain 
the spectrum of $H_J^{(L)}$. It can be obtained from the spectrum of an 
$U_q(sl(2))$ (for $q = e^{i\pi/3}$) symmetric quantum chain with $L$ sites (see \cite{NGRP}, eq.(2.12)) 
using the following rule: 
a) there is an unique ground-state of energy zero (if $L$ is odd, the ground 
state of the $U_q(sl(2))$ symmetric chain is doubly degenerate)   
b) the degeneracy of a non-zero energy level in $H_J^{(L)}$ is equal to 2/3 
the degeneracy of the same level in the $U_q(sl(2))$ symmetric chain.
Using this observation and the known results on the finite-size 
scaling of the spectra of the $U_q(sl(2))$ symmetric chain \cite{BS} 
for $q=e^{i\pi/3}$ one can 
easily derive the conformal properties of the Jacobstahl systems.

 We didn't try to generalize our observations to other spin chains.

Before closing our Letter, let us mention a fact observed independently 
by several people \cite{N-J-B}. The ground-state wavefunction of $H_{XXZ}^{(L)}$
given by 
eq.(\ref{eq3}) has positive integer coefficients. Positive coefficients 
occur if 
the ground-state wavefunction is associated with a Hamiltoninan which 
describes a stochastic process (the ground-sate energy has to be zero, 
which is the case). 
The ground-state wavefunction of a stochastic
process can be interpreted as a probability distribution function and
therefore has positive coefficients.
 One can show that in $H_J^{(L)}$, in the sector in which 
the ground-state is found,  one can define a stochastic process. If the fact 
that the coefficients are integer has a more profound explanation, remains 
to be seen.

Acknowledgements: This work was supported in part by the Brazilian agencies 
FAPESP and CNPq (Brazil), and VR acknowledges support from the EU network 
HPR-CT-2002-00325.

\end{document}